\documentclass[submission, Phys]{SciPost}

\usepackage{color}
\usepackage{float}
\usepackage{amssymb}

\usepackage{braket}
\usepackage[smalltableaux]{ytableau}
\usepackage{subcaption}
\begin{document}
\sloppy
\begin{center}{\Large \textbf{
Equilibration of quantum cat states}}\end{center}

\begin{center}
Tony Jin\textsuperscript{1} \textsuperscript{2}
\end{center}

\begin{center}
{\bf 1} DQMP, University of Geneva, 24 Quai Ernest-Ansermet, CH-1211 Geneva,
Switzerland
\\
{\bf 2} Laboratoire de Physique de l'\'Ecole Normale Sup\'erieure, CNRS, ENS $\&$ PSL University, Sorbonne Universit\'e, Universit\'e de Paris, 75005 Paris, France
\\
* zizhuo.jin@unige.ch
\end{center}

\begin{center}
\today
\end{center}


\section*{Abstract}
{\bf
We study the equilibration properties of isolated ergodic quantum
systems initially prepared in a \emph{cat state}, i.e a macroscopic
quantum superposition of states. Our main result consists in showing
that, even though decoherence is at work in the mean, there exists
a \emph{remnant }of the initial quantum coherences visible in the
strength of the fluctuations of the steady state. We back-up our analysis
with numerical results obtained on the XXX spin chain with a random
field along the z-axis in the ergodic regime and find good qualitative
and quantitative agreement with the theory. We also present and discuss
a framework where equilibrium quantities can be computed from general
statistical ensembles without relying on microscopic details about
the initial state, akin to the eigenstate thermalization hypothesis.
}

\vspace{10pt}
\noindent\rule{\textwidth}{1pt}
\tableofcontents\thispagestyle{fancy}
\noindent\rule{\textwidth}{1pt}
\vspace{10pt}

\section{Introduction}
\label{sec:intro}
Upon encountering the quantum statistical ensembles for the first
time, one is often struck by the strong similitude they share with
their classical counterpart. Indeed, quantum ensembles such as e.g
the Gibbs ensemble, appear like a mere transcription of classical
ones where one would have replaced the possible classical configurations
by the eigenstates of the Hamiltonian. An explanation dating back
to the early days of quantum mechanics \cite{Schrodinger,vonneumannergodicity}
is that, assuming ergodicity, the off-diagonal elements undergo a
dephasing that time-averages to zero given that the different frequencies
of the Hamiltonian are incommensurate. Thus, in the mean steady-state,
purely quantum mechanical features such as superposition of state and
entanglement are lost : this is a \emph{decoherence }effect. Furthermore,
the previous years have seen the development of a general framework
known as the \emph{eigenstate thermalization hypothesis }(ETH) which
explains the emergence of statistical ensembles from a given set of
assumptions on the spectral properties of the observables of the system
\cite{Deutschnoise,Deutsch_2018,ETH,Rigolreview} . The validity
or invalidity of the ETH has been tested numerically in a certain
number of studies \cite{GGERigol,Rigolreview,BreakdownofETHrigol,ETH1DboseGasUeda,NumericaltestETHkim}.

Therefore, one could legitimately ask what is the consequences of
having purely quantum features such as superposition of states and
entanglement in the initial state of the system on the final equilibrium
properties, if there are any at all? In this work, we intend to
prove that, even if on average information about the quantum coherence
of the initial state is lost at equilibrium, there is a remnant of
the latter visible in the \emph{fluctuations }around the stationary
state. This phenomenon was already seen in a model of stochastic fermionic
chain on a discrete lattice \cite{4BBJ,5BJ} and we provide here
the generalization of these results to \emph{any} ergodic quantum
system.

In the context of ETH, one important assumption is that the initial
states considered must have an energy comprised in a narrow energy
shell. This assumption is tightly bound with having initial states
which fulfill a \emph{cluster decomposition }\cite{weinberg_1995}\emph{
}constraint, i.e that the typical coherence length is small compared
to the size of the system. Within this hypothesis, the fluctuations
of the state around its average value scale like the inverse of the
dimension of the Hilbert space and are thus exponentially suppressed
as one increases the system size \cite{reviewthermalization}. 

\begin{figure}

\begin{center}
\includegraphics[scale=0.7]{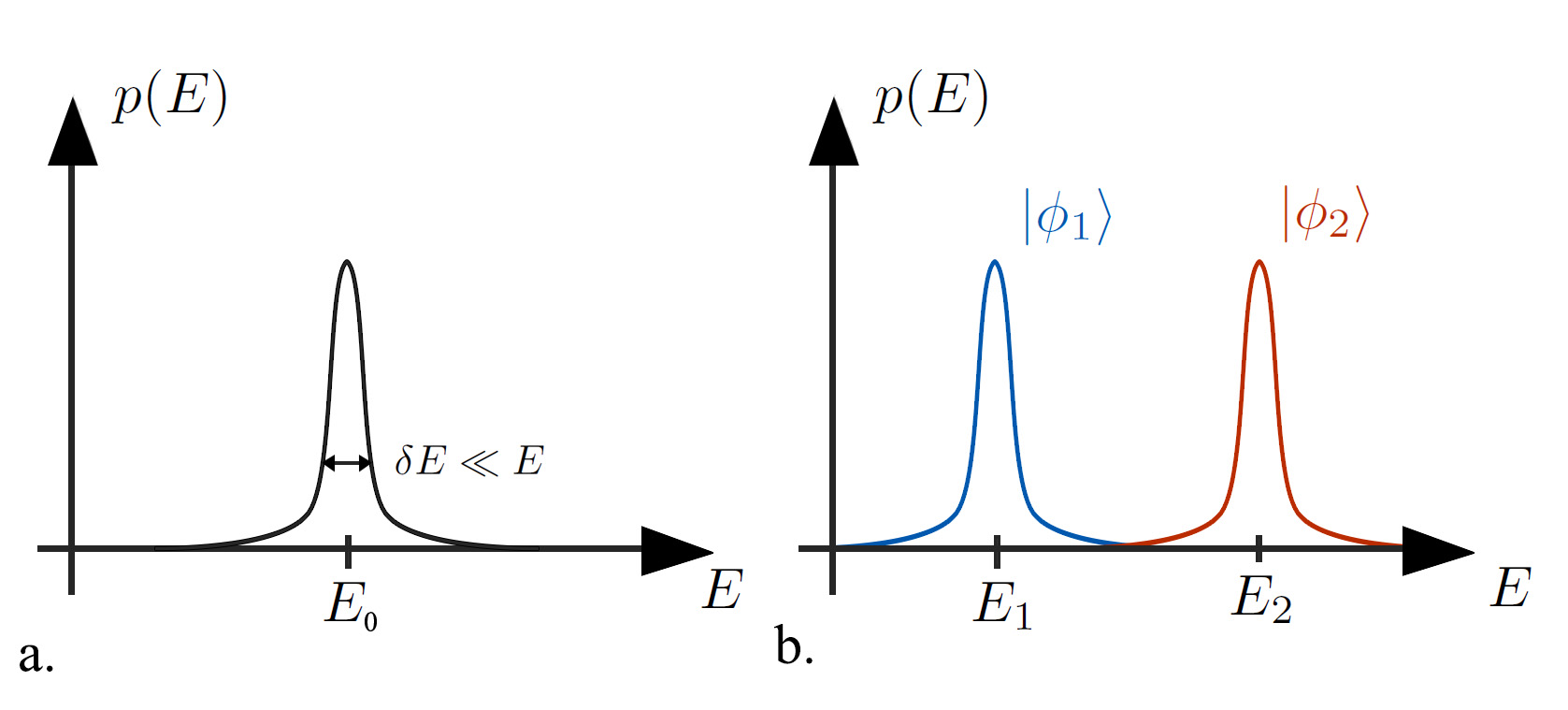}
\end{center}
\caption{a. Traditional situation where the set of initial states all belong
to a narrow energy window centered around $E_{0}$. b. Typical situation
we will consider in this paper where we have a cat state made of a
quantum superposition $1/\sqrt{2}(\ket{\Phi_{1}}+\ket{\Phi_{2}})$\label{fig:a.-Traditional-situation}}

\end{figure}

We are interested in situations where these hypothesis are relaxed,
i.e for which the initial state of the system can be a superposition
of states which have energies largely spread across the spectrum or
equivalently that entangle large part of the system together. This
typically happens for \emph{cat states }which are quantum superposition
of macroscopically distinct states -see fig.\ref{fig:a.-Traditional-situation}.
Cat states have attracted humongous interest from the physics community
in the recent years \cite{csZhao2004,csPhysRevX.8.021012,csMcConnell2015,csHaas180,lukinquantumsimulation,reinerblatt}
as their creation, stabilization and manipulation constitute key steps
towards quantum computing and simulation. Working in the Hamiltonian basis, we will see that such states
present non trivial, possibly non-local fluctuations of the off-diagonal
components in the steady-state that are fixed by the initial quantum
coherences.

This paper is organized as follows : First, we introduce our definition
for quantum ergodicity and following it, compute first and second
order correlation functions for elements of the density matrix. In
a second part, we show that these results are in qualitative and quantitative
agreement with numerical results obtained in a quantum ergodic spin
chain. We then discuss a more general framework where the equilibrium
ensembles don't depend on the fine structure of the initial states
and discuss connection with ETH. We finally end by some concluding
remarks and perspectives. 

\section{Ergodic hypothesis and equilibrium state}

In classical physics, ergodicity is the hypothesis that at long-times,
when the system reaches equilibrium, there is an equivalence between
the time average of quantities and an ensemble average over a \emph{microcanonical
}distribution. The microcanonical distribution stipulates that, at
fixed energy for an isolated system, the probability of all microscopic
configurations are equal. Physically, the equivalence between the
two averages comes from the assumption that at long-time the system
explores isotropically all the degrees of freedom available under
the constraint of fixed energy. 

This work is devoted to the formulation and study of a similar \emph{quantum
ergodic hypothesis} : Let $\rho_{0}$ be the density matrix containing
information about the initial conditions of the system. Given a set
of conserved observables, $\hat{H}_{1},\hat{H}_{2},\cdots\hat{H}_{n}$,
an accessible state is defined as a density matrix $\rho$ such that
there exists a unitary $U$ commuting with all the $\hat{H}_{i}$
and fulfilling $U\rho_{0}U^{\dagger}=\rho$. The quantum ergodic hypothesis
asserts that in the long-time equilibrium state, all accessible density
matrices have same probability weight or equivalently, that time average
of elements of $\rho_{t}$ is equivalent to ensemble average over
all possible unitary evolution $U$ that commutes with the conserved
quantities. For simplification, in this paper, we will consider a
unique conserved quantity $\hat{H}$ but generalization to a set of
mutually commuting observables is straightforward. 

Let us introduce some notations. We call ${\cal G}$ the group formed
by all the unitaries such that $[U,\hat{H}]=0$. We call $\ket{E_{i}^{\nu_{i}}}$
the eigenvector corresponding to energy $E_{i}$ for $\hat{H}$ where
$\nu_{i}$ is an index accounting for possible degeneracies. We will
call $d_{i}$ the dimension of the subspace associated to energy $E_{i}$.
Because of the commutativity of the $U$'s with $H$, the group ${\cal G}$
can be decomposed as a direct product of $SU(d_{i})$ : ${\cal G}=\times_{i}SU(d_{i})$.
Alternatively, this means that in the eigenbasis of the Hamitonian,
$U$ can be written in blocks indexed by $i$ with an element $U^{(i)}$
of $SU(d_{i})$ in each block. This constitutes a \emph{fundamental
representation }of ${\cal G}$. We also introduce the decomposition
of $\rho_{0}$ into different sectors $\rho_{0}^{(i,j)}$ defined
as $\rho_{0}^{(ij)}=\sum_{\nu_{i},\nu_{j}}{\rm tr}(\rho_{0}\ket{E_{i}^{\nu_{i}}}\bra{E_{j}^{\nu_{j}}})\ket{E_{i}^{\nu_{i}}}\bra{E_{j}^{\nu_{j}}}$
and $\rho_{0}=\sum_{i,j}\rho_{0}^{(i,j)}$. 
\[
\rho_{0}=\begin{pmatrix}\underbrace{\begin{pmatrix}\rho^{(11)}\end{pmatrix}}_{d_{1}\times d_{1}} & \underbrace{\begin{pmatrix} & \rho^{(12)}\end{pmatrix}}_{d_{2}\times d_{1}} & \cdots\\
\underbrace{\begin{pmatrix}\rho^{(21)}\end{pmatrix}}_{d_{1}\times d_{2}} & \underbrace{\begin{pmatrix} & \rho^{(2,2)}\end{pmatrix}}_{d_{2}\times d_{2}}\\
\vdots &  & \ddots
\end{pmatrix}.
\]

We will now illustrate how our quantum ergodic hypothesis allow to
compute equilibrium quantities. We begin by considering the \emph{average
} of $\rho$ denoted by $\mathbb{E}[\rho_{0}]$ with respect to the
ensemble average we just introduced. By definition : 
\begin{equation}
\mathbb{E}[\rho_{0}]=\int_{{\cal G}}d\eta(U)U\rho_{0}U^{\dagger}.\label{eq:moyenne}
\end{equation}
$\eta$ is the natural measure on ${\cal G}$, that is $d\eta(U)=\prod_{i}d\eta^{(i)}(U^{(i)})$
with $d\eta^{(i)}$ the Haar measure, i.e the unique invariant measure
on $SU(d_{i}$). The physical interpretation of this expression is
exactly the one we discussed before : The average evolution is given
by summing over all possible evolutions that preserve the spectrum
of the Hamiltonian with the probability weight of each of them distributed
uniformly with respect to the Haar measure. Making use of the decomposition
in sectors of $\rho_{0}$, we have : 
\begin{eqnarray}
\mathbb{E}[\rho_{0}] & = & \sum_{i}\mathbb{E}[\rho_{0}^{(i,i)}]+\sum_{i\neq j}\mathbb{E}[\rho_{0}^{(i,j)}]\\
 & = & \sum_{i}\int_{{\cal G}}d\eta^{(i)}(U^{(i)})U^{(i)}\rho_{0}^{(i,i)}U^{(i)\dagger}+\sum_{i\neq j}\int_{{\cal G}}d\eta^{(i)}(U^{(i)})d\eta^{(j)}(U^{(j)})U^{(i)}\rho_{0}^{(i,j)}U^{(j)\dagger}.
\end{eqnarray}
By the left invariance of the Haar measure we have that for $i\neq j$,
$\forall U^{(i)}\in SU(d_{i})$, $U^{(i)}\mathbb{E}[\rho_{0}^{(i,j)}]=\mathbb{E}[\rho_{0}^{(i,j)}]$
which is only true if $\mathbb{E}[\rho_{0}^{(i,j)}]=0$. For $i=j$,
Schur lemma tells us that $\mathbb{E}[\rho_{0}^{(i,j)}]$ must be
proportional to the identity. The proportionality coefficient is determined
by taking the trace so that we get : 
\begin{equation}
\mathbb{E}[\rho_{0}]=\sum_{i}\frac{1}{d_{i}}\mathbb{I}^{(ii)}{\rm tr}(\rho_{0}^{(ii)}).\label{eq:mean}
\end{equation}
Thus, in average, as one expects from decoherence, information about
''off-diagonal'' correlations between different energy sectors is
lost. However, we will see in what follows that there is actually
a remnant of the latter when one goes to higher order correlations.

Before going on, let's notice two extreme cases of interest : the first case is when all the energy levels are non-degenerate. Then, $\mathbb{E}[\rho_{0}]$ is just the diagonal ensemble, i.e the density matrix in which one has set all the off-diagonal components to zero. The second case is when there is only one energy sector that is the whole Hilbert space itself. We then have $\mathbb{E}[\rho_{0}]=\frac{1}{d}\mathbb{I}$. In this case, the density matrix states tells us all states with the same energy $E$ have the same probability weight, i.e it is the microcanonical ensemble. Fully-degenerate spectrum corresponds in general to chaotic or non-integrable systems, so one should expect that the diagonal ensemble describes accurately the steady state of such systems \cite{reviewthermalization}. However, in practice, we know that equilibrium states of isolated system are accurately described by the microcanonical ensemble which corresponds to the steady-state of a fully degenerate spectrum. To go from the first ensemble to the second is not a trivial task which requires additional assumptions. We will discuss this point in more details in the section \ref{Discussion}.

The second moment of elements of the density matrix is by definition
: 
\begin{align}
\mathbb{E}[\rho_{0}^{\otimes2}]=\int_{{\cal G}}d\eta(U)U^{\otimes2}\rho_{0}^{\otimes2}U^{\dagger\otimes2},
\end{align}
with $X^{\otimes n}\equiv\underbrace{X\otimes\cdots\otimes X}_{n\text{ times}}$.
This quantity can be computed by generalizing arguments used for the
mean. Again, it relies on the decomposition of $\rho_{0}^{\otimes2}$
into sectors $(\rho_{0}^{\otimes2})^{(i_{1},j_{1},i_{2},j_{2})}\equiv\sum_{\nu_{i_{1},}\nu_{i_{2},}\nu_{j_{1}},\nu_{j_{2}}}{\rm tr}(\rho_{0}^{\otimes2}\ket{E_{i_{1}}^{\nu_{i_{1}}},E_{i_{2}}^{\nu_{i_{2}}}}\bra{E_{j_{1}}^{\nu_{j_{1}}},E_{j_{2}}^{\nu_{j_{2}}}})\ket{E_{i_{1}}^{\nu_{i_{1}}},E_{i_{2}}^{\nu_{i_{2}}}}\bra{E_{j_{1}}^{\nu_{j_{1}}},E_{j_{2}}^{\nu_{j_{2}}}}$
and identifying the invariant objects under the action of $U\otimes U$.
We simply state the result and present the full derivation in app.\ref{sec:Second-order-fluctuations}:
\begin{eqnarray}
\mathbb{E}[\rho_{0}^{\otimes2}] & = & \sum_{i_{1}}\frac{1}{d_{i_{1}}(d_{i_{1}}+1)}(({\rm tr}(\rho_{0}^{(i_{1})}))^{2}+{\rm tr}((\rho_{0}^{(i_{1})})^{2}))\mathbb{I}_{(2)}^{(i_{1},i_{1},i_{1},i_{1})}\nonumber \\
 &  & +\frac{1}{d_{i_{1}}(d_{i_{1}}-1)}(({\rm tr}(\rho_{0}^{(i_{1})}))^{2}-{\rm tr}((\rho_{0}^{(i_{1})})^{2}))\mathbb{I}_{(1,1)}^{(i_{1},i_{1},i_{1},i_{1})}\label{eq:moment2}\\
 &  & +\sum_{i_{1}\neq i_{2}}\frac{1}{d_{i_{1}}d_{i_{2}}}(\mathrm{tr}(\rho_{0}^{(i_{1}i_{1})})\mathrm{tr}(\rho_{0}^{(i_{2}i_{2})})\mathbb{I}^{(i_{1}i_{2}i_{1}i_{2})}+\mathrm{tr}(\rho_{0}^{(i_{1}i_{2})}\rho_{0}^{(i_{2}i_{1})})\mathbb{I}^{(i_{1}i_{2}i_{2}i_{1})})\nonumber 
\end{eqnarray}
where the different identities are defined as : 
\begin{align}
\mathbb{I}_{(2)}^{(i_{1},i_{1},i_{1},i_{1})} & =\sum_{\nu_{i_{1}},\nu_{i_{1}}'}\frac{1}{4}(\ket{E_{i_{1}}^{\nu_{i_{1}}},E_{i_{1}}^{\nu_{i_{1}}^{'}}}+\ket{E_{i_{1}}^{\nu_{i_{1}}^{'}},E_{i_{1}}^{\nu_{i_{1}}}})(\bra{E_{i_{1}}^{\nu_{i_{1}}},E_{i_{1}}^{\nu_{i_{1}}^{'}}}+\bra{E_{i_{1}}^{\nu_{i_{1}}^{'}},E_{i_{1}}^{\nu_{i_{1}}}})\\
\mathbb{I}_{(1,1)}^{(i_{1},i_{1},i_{1},i_{1})} & =\sum_{\nu_{i_{1}}\neq\nu_{i_{1}}'}\frac{1}{4}(\ket{E_{i_{1}}^{\nu_{i_{1}}},E_{i_{1}}^{\nu_{i_{1}}^{'}}}-\ket{E_{i_{1}}^{\nu_{i_{1}}^{'}},E_{i_{1}}^{\nu_{i_{1}}}})(\bra{E_{i_{1}}^{\nu_{i_{1}}},E_{i_{1}}^{\nu_{i_{1}}^{'}}}-\bra{E_{i_{1}}^{\nu_{i_{1}}^{'}},E_{i_{1}}^{\nu_{i_{1}}}})\\
\mathbb{I}^{(i_{1},i_{2},i_{1},i_{2})} & =\sum_{\nu_{i_{1}},\nu_{i_{2}}}\ket{E_{i_{1}}^{\nu_{i_{1}}},E_{i_{2}}^{\nu_{i_{2}}}}\bra{E_{i_{1}}^{\nu_{i_{1}}},E_{i_{2}}^{\nu_{i_{2}}}},\\
\mathbb{I}^{(i_{1},i_{2},i_{2},i_{1})} & =\sum_{\nu_{i_{1}},\nu_{i_{2}}}\ket{E_{i_{1}}^{\nu_{i_{1}}},E_{i_{2}}^{\nu_{i_{2}}}}\bra{E_{i_{2}}^{\nu_{i_{2}}},E_{i_{1}}^{\nu_{i_{1}}}}.
\end{align}
The subscripts $(2)$ and $(1,1)$ refers respectively to the symmetric
and antisymmetric irreducible representations of $SU(d_{i})\otimes SU(d_{i})$.
The important point is that contrary to (\ref{eq:mean}), (\ref{eq:moment2})
contains information about quantum superposition of states both in
the case where they belong to the same sector (line 1 of (\ref{eq:moment2}))
but also when the superposition involves states belonging to different
sectors (line 3 of (\ref{eq:moment2})). Thus, two initial states
having the same diagonal elements may relax to the same density matrix
in average but present differences at the level of fluctuating quantities,
providing a signature of the presence or the absence of initial quantum
coherences.

One can carry this procedure to get access to higher moments but their
explicit expression becomes more and more involved. We present the
general formula in the app.\ref{sec:General-formula-at}.

We will now illustrate these ideas on a concrete numerical example. 

\section{Numerical results on the XXX spin chain with random field in the ergodic regime}

We test the predictive power of our model on the XXX model with random
local fields :
\begin{align}
\hat{H}=\sum_{j=0}^{L-2}J\vec{\sigma}_{j}\cdot\vec{\sigma}_{j+1}+\sum_{j=0}^{L-1}h_{j}\sigma_{j}^{z}
\end{align}
with $L$ the lattice size and $\sigma_{j}^{a}$ the usual Pauli matrices.
The boundaries are open. The $h_{j}$ are independent random variables
distributed uniformly in an interval $[-h,h]$. The transition from
an ergodic to a localized regime of this model has been studied in
\cite{mobilityedgePhysRevB.91.081103} and characterized by the spectral
properties of $\hat{H}$. A quantity of particular interest is the
mean ratio of consecutive level spacings known to be close to the one of the Wigner
distribution ($\approx0.53$) in the ergodic regime and to the one
of the Poisson distribution ($\approx0.38$) in the localized regime.
For $J=1$ and lattice size ranging from $11$ to $22$, it has bee
shown in \cite{mobilityedgePhysRevB.91.081103} that the transition
between the two regimes occurred for $h\approx2.5$. Since we are
interested in the ergodic regime we will fix the value of $h$ to
$1$. We work in the minimal magnetization sector, i.e $0$ for $L$
even and $1$ for $L$ odd. 

Let $\hat{O}$ be an observable. We will compute the time-evolution
of $O(t)\equiv{\rm tr}(\rho_{t}\hat{O})$ by using exact diagonalization
methods \cite{qutipJOHANSSON20121760,EDSciPostPhys.2.1.003,ED10.21468/SciPostPhys.7.2.020}.
We denote the time-average by $\mathbb{E}_{t}[\bullet]\equiv\lim_{T\to\infty}\frac{1}{T}\int_{t=0}^{T}\bullet dt$
and will be interested in first and second order correlation functions
$\mathbb{E}_{t}[O(t)]$, $\mathbb{E}_{t}[O(t)O'(t)]$. Our point will
be to show that the time average $\mathbb{E}_{t}[\bullet]$ is equivalent
to the previously introduced ensemble average over possible unitary
evolutions $\mathbb{E}[\bullet]$. 

We will study a \emph{quench }situation in which the initial state
is expressed in terms of eigenvalues of the Hamiltonian $\hat{H}$
from which we remove the XXX coupling between sites $L/2-1$ and $L/2$
(suppose $L$ even for simplicity), i.e $\hat{H}_{0}=\hat{H}_{{\rm L}}+\hat{H}_{{\rm R}}$
with $\hat{H}_{{\rm L}}\equiv\sum_{j=0}^{L/2-2}J\vec{\sigma}_{j}\cdot\vec{\sigma}_{j+1}+\sum_{j=0}^{L/2-1}h_{j}\sigma_{j}^{z}$
and $\hat{H}_{{\rm R}}\equiv\sum_{j=L/2}^{L-2}J\vec{\sigma}_{j}\cdot\vec{\sigma}_{j+1}+\sum_{j=L/2}^{L-1}h_{j}\sigma_{j}^{z}$.
The initial states chosen this way have well-defined energies $E_{{\rm R}}$
and ${\rm E}_{{\rm L}}$ and will be denoted $\ket{E_{{\rm R}},E_{{\rm L}}}$.
At time $t=0$, we switch the Hamiltonian from $\hat{H}_{0}$ to $\hat{H}$,
so that the system is now in an out-of-equilibrium situation. 

To illustrate the importance of the presence or absence of initial
quantum coherences in the steady state of the system, we propose to
study two different set of initial conditions. They will be both indistinguishable
from the point of view of their mean energy but they will encode for
different off-diagonal quantum coherences which effect will be visible
in the equilibrium fluctuations of the system. Let $E_{{\rm min}}$,
$E_{{\rm max}}$ be the minimum and maximum energy of the spectrum
and $\ket{\Phi_{1}}\equiv\ket{E_{{\rm R}}^{1},E_{{\rm L}}^{1}}$,
$\ket{\Phi_{2}}=\ket{E_{{\rm R}}^{2},E_{{\rm L}}^{2}}$ such that
$E_{{\rm R}}^{1}+E_{{\rm L}}^{1}$ is close to $E_{{\rm min}}$ and
$E_{{\rm R}}^{2}+E_{{\rm L}}^{2}$ is close to $E_{{\rm max}}$. The
decomposition of these states in the eigenbasis of $\hat{H}$ are
shown in the app.\ref{sec:More-details-on}.

In the \emph{protocol I, }corresponding to a \emph{cat state }made
of a quantum superposition of two states with <<macroscopically>>
distinct energies $E_{{\rm R}}^{1}+E_{{\rm L}}^{1}$ and $E_{{\rm R}}^{2}+E_{{\rm L}}^{2}$\emph{,
}the initial state is chosen to be : 
\begin{align}
\ket{\psi_{0}^{{\rm I}}}=\frac{1}{\sqrt{2}}(\ket{\Phi_{1}}+\ket{\Phi_{2}}).
\end{align}
In the \emph{protocol II}, corresponding to a \emph{mixed
state}, the initial state is described by the density matrix : 
\begin{align}
\rho_{0}^{{\rm II}}=\frac{1}{2}(\ket{\Phi_{1}}\bra{\Phi_{1}}+\ket{\Phi_{2}}\bra{\Phi_{2}}).
\end{align}
In both protocols, the reduced density matrices on ${\rm R}$ and
${\rm L}$ are the same. 

We compute the time-evolution of two observables : $H_{{\rm R}}(t)\equiv{\rm tr}(\rho_{t}\hat{H}_{{\rm R}})$
and $Q(t)={\rm tr}(\rho_{t}\hat{Q})$, $\hat{Q}\equiv\ket{\Phi_{1}}\bra{\Phi_{2}}+\ket{\Phi_{2}}\bra{\Phi_{1}}$. Note that $Q$ is non-local, in the sense that it has non zero support on the whole physical space.
From formula (\ref{eq:mean},\ref{eq:moment2}) we can deduce the
predictions for first and second moments of these quantities given
by ensemble averages in both protocols. Importantly we have that :
\begin{align}
\mathbb{E}^{{\rm I}}[H_{{\rm R}}]=\mathbb{E}^{{\rm II}}[H_{{\rm R}}],\quad & \mathbb{E}^{{\rm I}}[H_{{\rm R}}^{2}]=\mathbb{E}^{{\rm II}}[H_{{\rm R}}^{2}],\\
\mathbb{E}^{{\rm I}}[Q]=\mathbb{E}_{}^{{\rm II}}[Q]=0,\quad & \mathbb{E}^{{\rm I}}[|Q|^{2}]\neq\mathbb{E}^{{\rm II}}[|Q|^{2}],
\end{align}
meaning that the two protocols can be distinguished by looking at
the fluctuations of $\hat{Q}$. Qualitatively, this comes from the
fact that the observable $\hat{Q}$ has non zero projection on off-diagonal
elements of the energy basis of $\hat{H}$. The fluctuations of the
latter is precisely what characterizes the difference between the
equilibrium state of protocol I and II. The computations and detailed
expressions of these quantities are provided in app.\ref{sec:More-details-on}.

\begin{figure}
\begin{minipage}[t]{0.49\columnwidth}%
\begin{center}
\includegraphics[scale=0.30]{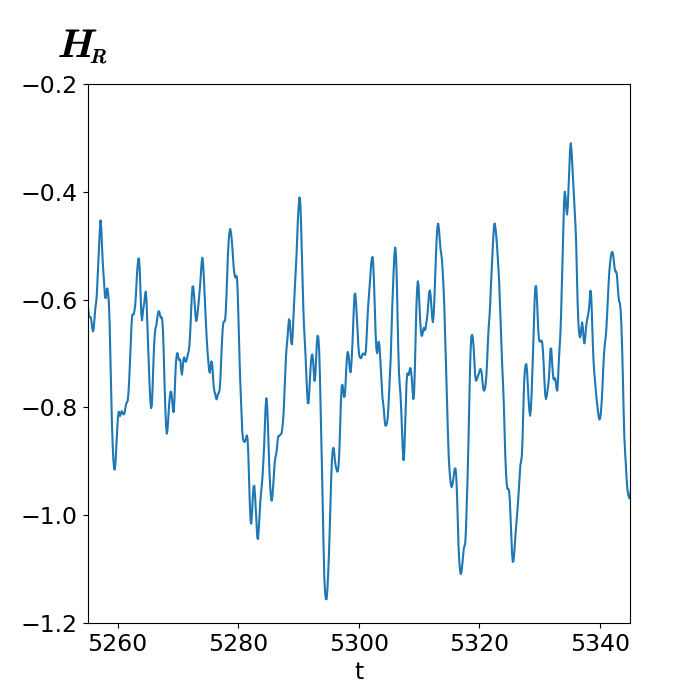}%
\end{center}
\end{minipage}%
\begin{minipage}[t]{0.49\columnwidth}%
\begin{center}
\includegraphics[scale=0.30]{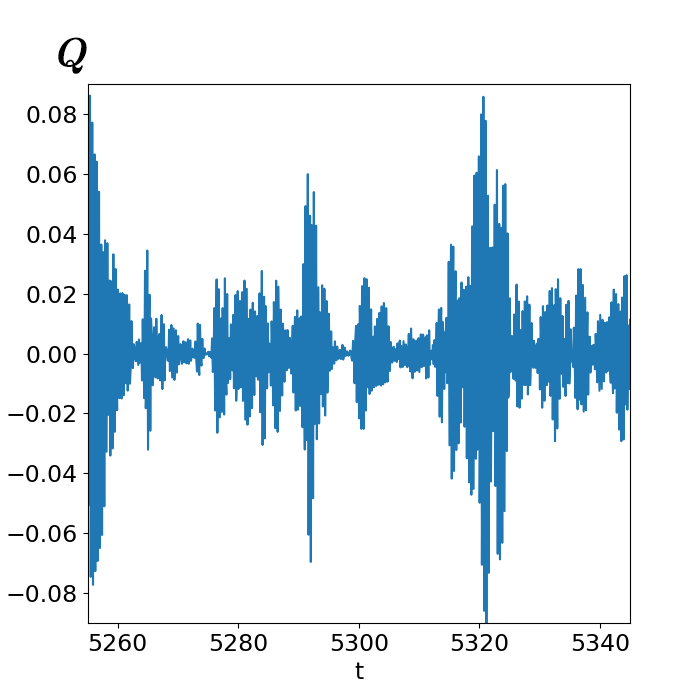}%
\end{center}
\end{minipage}

\begin{minipage}[t]{0.49\columnwidth}%
\begin{center}
\includegraphics[scale=0.30]{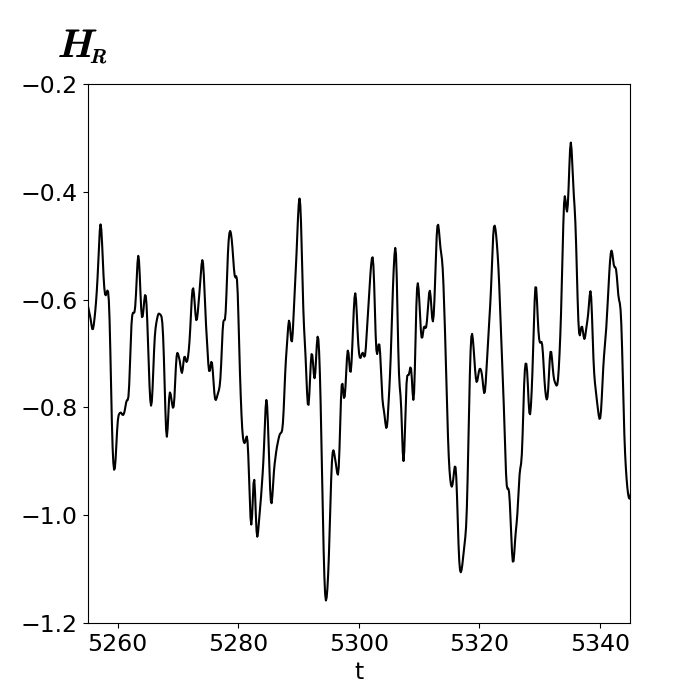}%
\end{center}
\end{minipage}%
\begin{minipage}[t]{0.49\columnwidth}%
\begin{center}
\includegraphics[scale=0.30]{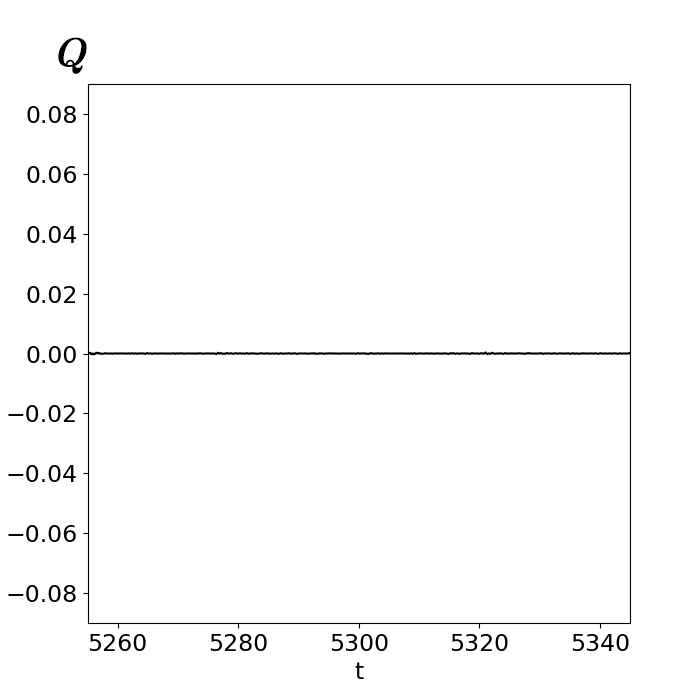}%
\end{center}
\end{minipage}

\caption{Long-time evolution of the different quantities considered. In \textbf{\textcolor{blue}{\emph{blue}}}
are the plots corresponding to the \textbf{\textcolor{blue}{\emph{cat
state }}}\textbf{\textcolor{blue}{(protocol I)}} while in \textbf{\emph{black}}
are the plots for the \textbf{\emph{mixed state}} \textbf{(protocol
II)}. We see no qualitative difference for $H_{{\rm R}}$ while the
fluctuations of $Q$ around the mean are suppressed for the second
protocol, as the consequence of the absence of initial quantum coherences. The blue-shaded region in the top-right panel is the consequence of oscillations occurring on a much shorter time scale.\label{fig:Long-time-evolution-of}}
\end{figure}

We show in fig.\ref{fig:Long-time-evolution-of}, the time-evolution
of $H_{{\rm R}}(t)$ and $Q(t)$ in both protocols for a given realization
of the disorder. We can clearly see that $H_{{\rm R}}(t)$ is independent
of the protocol, contrarily to $Q(t)$. The predicted value for the
different quantities is also in quantitative arguments with the simulations
(see tab.\ref{tab:}).
Details on how these values and the confidence intervals are obtained are given in app.\ref{sec:More-details-on}. 

\begin{table}
\begin{center}
\begin{tabular}{|c|c|c|c|}
\hline 
Theory & $\mathbb{E}[H_{R}]$ & $\sigma_{H_{R}}$ & $\sigma_{Q}$\tabularnewline
\hline 
Cat state & $-7.13*10^{-1}$ & $1.51*10^{-1}$ & $2.09*10^{-2}$\tabularnewline
\hline 
Mixed state & $-7.14*10^{-1}$ & $1.51*10^{-1}$ & $\approx0$\tabularnewline
\hline 
\end{tabular}
\end{center}
\bigskip{}
\begin{center}
\begin{tabular}{|c|c|c|c|}
\hline 
Numeric & $\mathbb{E}[H_{R}]$ & $\sigma_{H_{R}}$ & $\sigma_{Q}$\tabularnewline
\hline 
Cat state & $-7.14*10^{-1}\pm7.8*10^{-4}$ & $1.52*10^{-1}\pm0.3*10^{-2}$ & $2.11*10^{-2}\pm0.2*10^{-2}$\tabularnewline
\hline 
Mixed state & $-7.14*10^{-1}\pm7.8*10^{-4}$ & $1.52*10^{-1}\pm0.3*10^{-2}$ & $5.83*10^{-5}\pm1.0*10^{-6}$\tabularnewline
\hline 
\end{tabular}\caption{Comparison between theoretical and numerical values for the mean and
standard deviation of different quantities of interest. \label{tab:}}
\end{center}
\end{table}

Let us add a remark here. In general because of its high degree of non-locality, it is not expected that $Q$ might be a suitable observable for experimental measurements. But similar qualitative statements about fluctuations should apply for any observables that couple the different energy sectors. For instance, as suggested at the end of \cite{GullansHuse}, one could imagine doing an interference experiment between two parts of the system far part and look at the fluctuations of the pattern.
\section{Discussion}\label{Discussion}

So far, we were only concerned about the \emph{equilibrium state }of
the system and haven't gone into the \emph{thermalization }properties.
Thermalization is stronger as it implies that the steady-state properties
of the system can be described by one of the canonical ensemble of
thermodynamics. In this section, we informally discuss possible links
between the theory presented and the Eigenstate Thermalization Hypothesis
(ETH). The ETH conjectures that for any initial state prepared as
mixture of eigenstates of the total Hamiltonian with energies in a
narrow window $[E-\delta E,E+\delta E]$, the matrix elements of observables
in the energy eigenbasis is given by $O_{mn}=O(\overline{E})\delta_{mn}+e^{-S(\overline{E})/2}f_{O}(\overline{E},\omega)R_{mn}$
with $\overline{E}\equiv(E_{m}+E_{n})/2$, $\omega\equiv E_{n}-E_{m}$
and $S(\overline{E})$ the entropy. It is assumed that $O(\overline{E})$
and $f_{O}(\overline{E},\omega)$ are \emph{smooth }function of their
arguments and that $R_{mn}$ is a random variable with zero mean and
unit variance.

In the ETH, the role of the initial state is restricted to fixing the energy scales $E$ and $\Delta E$. The important remark is that, for the cat states, there is no notion of ''narrow window'' around a given energy anymore, hence we don't expect the ETH to apply. One illustration of that is the fact that off-diagonal correlations are not exponentially suppressed for cat states.

However the great advantage of the ETH is to explain why one can forget
about all microscopic details contained in the diagonal ensemble and
instead work with the microcanonical ensemble $\rho_{{\rm m}}\equiv\frac{1}{d}\mathbb{I}$.
It would be great to have an equivalent statement here. A possible
way for obtaining such simplification in our case already discussed
in \cite{6BBJPhysRevE.101.012115} would be the following : We can
suppose that the ''actual'' group whose action leaves invariant
the stationary state is not given by the set of all unitaries that
commutes with $H$ but rather with an Hamiltonian $H'=H+\delta H$
with $\delta H$ a small perturbation which "mixes" the different
energy sectors separated by energy $\approx\delta H$. The microcanonical
ensemble is recovered in the case where the spectrum of $H'$ is fully
degenerate in the energy window of interest $[E-\delta E,E+\delta E]$.
Indeed, in that case, from (\ref{eq:mean}) we see that the average
density matrix is the microcanonical one : $\mathbb{E}[\rho_{0}]=\rho_{{\rm m}}$
and the second moment is $\mathbb{E}[\rho_{0}\otimes\rho_{0}]=\frac{2}{d(d+1)}\mathbb{I}_{\{2\}}$
(for simplicity, we suppose that the initial state is a pure state).
For the case where the initial state has two peaks $E_{1}$ and $E_{2}$
in its energy spectrum as shown in fig.\ref{fig:a.-Traditional-situation},
we can conjecture that $H'$ is such that it mixes energies in the
interval ${\rm I}_{1}=[E_{1}-\delta E,E_{1}+\delta E]$ and ${\rm I}_{2}=[E_{2}-\delta E,E_{2}+\delta E]$
around $E_{1}$ and $E_{2}$ but not altogether so that the average
density matrix is given by : $\mathbb{E}[\rho_{0}]=\frac{{\rm tr}(\rho_{0}\mathbb{I}_{1})}{d_{1}}\mathbb{I}^{(1)}+\frac{{\rm tr}(\rho_{0}\mathbb{I}_{2})}{d_{2}}\mathbb{I}^{(2)}$
with $\mathbb{I}^{(i)}=\sum_{E\in{\rm I}_{i}}\ket{E}\bra{E}$ and
$d_{i}={\rm tr}(\mathbb{I}^{(i)})$. This has the simple interpretation
that, on average at equilibrium, the mean density matrix is a statistical
mixture of a state at energy $E_{1}$ and a state at energy $E_{2}$.
The only information retained from the initial state is the weights
corresponding to each energy sector. Similarly, a direct application
of (\ref{eq:moment2}) shows that at the second order, the information
about the connected correlations between the two energy sectors is
contained in a compact way in ${\rm tr}(\rho_{0}\otimes\rho_{0}\hspace{1em}\mathbb{I}^{(1,2,2,1)})$
with $\mathbb{I}^{(1,2,2,1)}=\sum_{E\in{\rm I}_{1},E'\in{\rm I}_{2}}\ket{E,E'}\bra{E',E}$.
Thus, one would not need fine information about the initial state
to describe the equilibrium properties of the system. Of course, as
with ETH, the range of applicability of these hypothesis is for now
rather elusive and needs to be determined via careful numerical or
experimental studies. We wish to report more on that in future studies.

\section{Conclusion}

We presented a theoretical framework enabling one to compute equilibrium
properties of isolated quantum systems upon an assumption of \emph{quantum
ergodicity }which postulates that time-averages are equivalent to
unitary ensemble averages in the stationary state.\emph{ }We brought
specific attention to the relaxation of \emph{cat states, }i.e quantum
states which are a superposition of two macroscopically distinct states.
We showed both analytically and numerically that a remnant of the
initial quantum coherence was visible in the fluctuations around average
quantities in the steady state whose amplitudes can be computed exactly.
In the last part, we sketched a possible framework describing the
equilibrium fluctuations in terms of statistical ensembles that do
not require full knowledge of the microscopic details of the initial
state.

In non-integrable or integrable systems a subject of debate of the
previous decades has been to determine which conserved quantities
were relevant to describe the thermal ensembles determining the local
equilibrium properties. The question is of particular relevance for
integrable systems since they comprise in principle a macroscopic
number of conserved quantities \cite{EsslerreviewGGE}. It is now
commonly accepted that one should only consider local (or quasilocal
\cite{quasilocalcharges}) quantities to characterize such ensembles.
However, our study stipulates that these ensembles no longer suffice
when one looks at the equilibrium fluctuations of the system. There,
additional information about possibly \emph{non-local }conserved quantities
are required. 

Another important affirmation of ETH concerns the notion of \emph{typicality
}\cite{typicalityPhysRevLett.80.1373,typicalityPhysRevLett.96.050403}\emph{.
}Typicality states that for all pure states that are random superpositions of eigenstates of the energy window,
few-body operators have thermal distributions in the thermodynamic
limit. It would be very interesting to test whether some notion of
typicality remains in our case, i.e that the fluctuations of -possibly
\emph{non-local- }few-body observables are described by typical distributions
in the thermodynamic limit starting from \emph{any }superposition
of states belonging to the macroscopically different energy sectors. 

Another interesting point is that the equilibrium formulae (\ref{eq:mean},\ref{eq:moment2})
can in principle be applied to non chaotic or integrable system. Of
course, there is no reason for the ergodic property to be fulfilled
anymore so there is no guarantee that they provide the right predictions.
For instance, for finite-size integrable systems, there might be long-lived oscillations that prevents the system
from equilibrating \cite{XYmodel,BlochBECrevival}. However one should
remark that the various symmetries of the system that lead to ergodicity
breaking \emph{are }accounted for in the structure of the group. It
would therefore be interesting to see whether this information about
degeneracies is enough to predict quantitatively the time-averaged
and amplitudes of oscillating quantities and, if not, what ingredient
needs to be added. 

\section*{Acknowledgements}
This work wouldn't have been possible without past and present collaborations
and discussions with D. Bernard and M. Bauer. I greatly benefited
from crucial feedback from B. Appfel, P. Caucal, D. Martin and M.
Rieu. I am also grateful for the work done by A. Gallin and T. Orlovic.
The numerics were performed using the QuSpin and QutiP Python packages.

\paragraph{Funding information}

I acknowledge support from the French École doctorale 564 and the Swiss National
Science Foundation, Division II.

\begin{appendix}

\section{Second order fluctuations\label{sec:Second-order-fluctuations}}
In this section we show how to compute formula (\ref{eq:moment2})
from the main text , i.e we want to compute : 
\[
\mathbb{E}[\rho_{0}^{\otimes2}]=\int_{{\cal G}}d\eta(U)U^{\otimes2}\rho_{0}^{\otimes2}U^{\dagger\otimes2}.
\]
In essence, the calculations will rely on the same mechanics than
for order $1$ with some twists. Once again, we define the decomposition
of $\rho_{0}^{\otimes2}$ into different sectors $(\rho_{0}^{\otimes2})^{(i_{1},j_{1},i_{2},j_{2})}$
as follows : 
\[
(\rho_{0}^{\otimes2})^{(i_{1},j_{1},i_{2},j_{2})}=\sum_{\nu_{i_{1}},\nu_{i_{2}},\nu_{j_{1}},\nu_{j_{2}}}{\rm tr}(\rho_{0}^{\otimes2}\ket{E_{i_{1}}^{\nu_{i_{1}}},E_{i_{2}}^{\nu_{i_{2}}}}\bra{E_{j_{1}}^{\nu_{j_{1}}},E_{j_{2}}^{\nu_{j_{2}}}})\ket{E_{i_{1}}^{\nu_{i_{1}}},E_{i_{2}}^{\nu_{i_{2}}}}\bra{E_{j_{1}}^{\nu_{j_{1}}},E_{j_{2}}^{\nu_{j_{2}}}}.
\]
The average of a block is given by : 
\[
\mathbb{E}[(\rho_{0}^{\otimes2})^{(i_{1},j_{1},i_{2},j_{2})}]=\int_{{\cal G}}d\eta(U)U^{(i_{1})}\otimes U^{(i_{2})}(\rho_{0}^{\otimes2})^{(i_{1},j_{1},i_{2},j_{2})}U^{\dagger(j_{1})}\otimes U^{\dagger(j_{2})}.
\]
We first prove that $\mathbb{E}[(\rho_{0}^{\otimes2})^{(i_{1},j_{1},i_{2},j_{2})}]$
is null except if the tuple $\{j_{1},j_{2}\}$ is a permutation of
$\{i_{1},i_{2}\}$. Indeed, suppose it's not the case : then, there
exists a $k$ such that $\forall k'$, $i_{k}\neq j_{k'}$. For definiteness,
say $k=1$. From the left invariance of the Haar measure, we then
have $\forall V^{(i_{1})}\in SU(d_{i_{1}})$ that : 
\begin{eqnarray*}
V^{(i_{1})}\otimes\mathbb{I}^{(i_{2})}\mathbb{E}[(\rho_{0}^{\otimes2})^{(i_{1},j_{1},i_{2},j_{2})}] & = & \mathbb{E}[(\rho_{0}^{\otimes2})^{(i_{1},j_{1},i_{2},j_{2})}]\\
(V^{(i_{1})}\otimes\mathbb{I}^{(i_{2})}-\mathbb{I}^{(i_{1})}\otimes\mathbb{I}^{(i_{2})})\mathbb{E}[(\rho_{0}^{\otimes2})^{(i_{1},j_{1},i_{2},j_{2})}] & = & 0
\end{eqnarray*}
which implies $\mathbb{E}[(\rho_{0}^{\otimes2})^{(i_{1},j_{1},i_{2},j_{2})}]=0$. 

We thus learn the important fact that $\{j_{1},j_{2}\}$ must be a
permutation of $\{i_{1},i_{2}\}$ for the average of the block to
be non zero.

There are three possible cases that we will examine separately : 
\begin{itemize}
\item[I] $i_{1}=i_{2}=j_{1}=j_{2}$,
\item[II] $i_{1}\neq i_{2}$, $i_{1}=j_{1}$, $i_{2}=j_{2}$ corresponding
to the permutation $\sigma\in\mathfrak{S}_{2}:\{1,2\}\to\{1,2\},$
\item[III] $i_{1}\neq i_{2}$, $i_{1}=j_{2}$, $i_{2}=j_{1}$ corresponding
to the permutation $\sigma\in\mathfrak{S}_{2}:\{1,2\}\to\{2,1\}$.
\end{itemize}

\paragraph{Case I : }

\[
[(\rho_{0}^{\otimes2})^{(i_{1}i_{1}i_{1}i_{1})}]=\int d\eta(U)U^{(i_{1})}\otimes U^{(i_{1})}(\rho_{0}\otimes\rho_{0})^{(i_{1}i_{1}i_{1}i_{1})}U^{\dagger(i_{1})}\otimes U^{\dagger(i_{1})}.
\]
Let $D^{(i)}$ be the fundamental representation of $SU(d_{i})$.
The tensor product representation $D^{(i)}\otimes D^{(i)}$ admits
a decomposition onto irreducible representations indexed by Young
diagrams. We denote by $\{2\}$ the possible partitions of $2$, i.e
$(2)$ and $(1,1)$. Following the usual convention for indexation
of irreducible representations of the unitary group by Young tableaux,
$D^{(2)(i)}$ corresponds to $\ydiagram{2}$ and the tensor representation
that is symmetric under permutation of two indices while $D^{(1,1)(i)}$
corresponds to $\ydiagram{1,1}$ and denotes the antisymmetric representation.
We have \cite{liealgebrainparticlephysicsGeorgi2018} : 
\[
D^{(i)}\otimes D^{(i)}=D^{(2)(i)}\oplus D^{(1,1)(i)}.
\]

The representation $D^{(2)(i)}$ preserves the symmetric eigenbasis
made of $\frac{d_{i}(d_{i}+1)}{2}$ elements $\left|E_{i},(2),\nu_{i},\nu'_{i}\right\rangle \equiv\frac{1}{\sqrt{2}}(\ket{E_{i}^{\nu_{i}},E_{i}^{\nu_{i}'}}+\ket{E_{i}^{\nu_{i}'},E_{i}^{\nu_{i}}})$
for $\nu_{i}\neq\nu'_{i}$ and $\left|E_{i},(2),\nu_{i},\nu{}_{i}\right\rangle =\left|E_{i}^{\nu_{i}},E_{i}^{\nu_{i}}\right\rangle $
while the representation $D^{(1,1)(i)}$ preserves the antisymmetric
eigenbasis made of $\frac{d_{i}(d_{i}-1)}{2}$ elements $\left|E_{i},(1,1),\nu_{i},\nu'_{i}\right\rangle =\frac{1}{\sqrt{2}}(\ket{E_{i}^{\nu_{i}},E_{i}^{\nu_{i}'}}-\ket{E_{i}^{\nu_{i}'},E_{i}^{\nu_{i}}})$
for $\nu_{i}\neq\nu_{i}'$.

One can further block-decompose $(\rho_{0}\otimes\rho_{0})^{(i_{1},i_{1},i_{1},i_{1})}$
according to these basis. Introducing :
\begin{align*}
 & (\rho_{0}^{\otimes2})^{(i_{1},i_{1},i_{1},i_{1}),(y_{1},y_{2})}\\
 & \equiv\sum_{\nu_{i},\nu'_{i},\mu_{i},\mu'_{i}}{\rm tr}((\rho_{0}^{\otimes2})^{(i_{1},i_{1},i_{1},i_{1})}\left|E_{i_{1}},(y_{1}),\nu_{i_{1}},\nu'_{i_{1}}\right\rangle \left\langle E_{i_{1}},(y_{2}),\mu_{i_{1}},\mu'_{i_{1}}\right|)\left|E_{i_{1}},(y_{1}),\nu_{i_{1}},\nu'_{i_{1}}\right\rangle \left\langle E_{i_{1}},(y_{2}),\mu_{i_{1}},\mu'_{i_{1}}\right|
\end{align*}
with $y_{1},y_{2}\in\{2\}$. 

By Schur lemma, we then have than the only non-zero block components
of $\mathbb{E}[(\rho_{0}^{\otimes2})^{(i_{1}i_{1}i_{1}i_{1})}]$ are
the diagonal ones, i.e $\mathbb{E}[(\rho_{0}^{\otimes2})^{(i_{1}i_{1}i_{1}i_{1}),((2),(2))}]$
and $\mathbb{E}[(\rho_{0}^{\otimes2})^{(i_{1}i_{1}i_{1}i_{1}),((1,1),(1,1))}]$
and these blocks are proportional to the identity : 
\begin{align*}
\mathbb{E}[(\rho_{0}^{\otimes2})^{(i_{1}i_{1}i_{1}i_{1}),((2),(2))}] & \propto\mathbb{I}_{(2)}^{(i_{1},i_{1},i_{1},i_{1})},\\
\mathbb{E}[(\rho_{0}^{\otimes2})^{(i_{1}i_{1}i_{1}i_{1}),((1,1),(1,1))}] & \propto\mathbb{I}_{(1,1)}^{(i_{1},i_{1},i_{1},i_{1})}.
\end{align*}
Where $\mathbb{I}_{\{2\}}^{(i_{1},i_{1},i_{1},i_{1})}$ is the identity
matrix associated to the representation $D^{\{2\}(i_{1})}$. The proportionality
coefficient is determined by taking the trace. Explicitly, we have
: 
\begin{eqnarray*}
\mathbb{E}[(\rho_{0}^{\otimes2})^{(i_{1}i_{1}i_{1}i_{1}),((2),(2))}] & = & \frac{2}{d_{i_{1}}(d_{i_{1}}+1)}{\rm tr}((\rho_{0}^{\otimes2})\mathbb{I}_{(2)}^{(i_{1},i_{1},i_{1},i_{1})})\mathbb{I}_{(2)}^{(i_{1},i_{1},i_{1},i_{1})}\\
 & = & \frac{1}{d_{i_{1}}(d_{i_{1}}+1)}(({\rm tr}(\rho_{0}^{(i_{1})}))^{2}+{\rm tr}((\rho_{0}^{(i_{1})})^{2}))\mathbb{I}_{(2)}^{(i_{1},i_{1},i_{1},i_{1})}\\
\mathbb{E}[(\rho_{0}^{\otimes2})^{(i_{1}i_{1}i_{1}i_{1}),((1,1),(1,1))}] & = & \frac{2}{d_{i}(d_{i}-1)}{\rm tr}((\rho_{0}^{\otimes2})\mathbb{I}_{(1,1)}^{(i_{1},i_{1},i_{1},i_{1})})\mathbb{I}_{(1,1)}^{(i_{1},i_{1},i_{1},i_{1})}.\\
 & = & \frac{1}{d_{i}(d_{i}-1)}(({\rm tr}(\rho_{0}^{(i_{1})}))^{2}-{\rm tr}((\rho_{0}^{(i_{1})})^{2}))\mathbb{I}_{(1,1)}^{(i_{1},i_{1},i_{1},i_{1})}
\end{eqnarray*}

\paragraph{Case II : }

We look at : 

\[
\mathbb{E}[(\rho_{0}^{\otimes2})^{(i_{1}i_{2}i_{1}i_{2})}]=\int d\eta(U)U^{(i_{1})}\otimes U^{(i_{2})}(\rho_{0}^{\otimes2})^{(i_{1}i_{2}i_{1}i_{2})}U^{\dagger(i_{1})}\otimes U^{\dagger(i_{2})}
\]
for $i_{1}\neq i_{2}$.

Since $D^{(i_{1})}$ and $D^{(i_{2})}$ are two irreducible representations
of $SU(d_{i_{1}})$ and $SU(d_{i_{2}})$, the tensor product $D^{(i_{1})}\otimes D^{(i_{2})}$
is an irreducible representation of $SU(d_{i_{1}})\times SU(d_{i_{2}})$. 

The proof comes from \emph{Schur orthogonality relation }:

Let $D_{1}$ and $D_{2}$ be two irreducible representations over
vector spaces $V_{1}$ and $V_{2}$ : $D_{1}:G1\to{\rm End}(V1)$,
$D_{2}:G2\to{\rm End}(V2)$.

If $G1$ and $G2$ are finite, we show that the tensor product representation
$D_{\otimes}=D_{1}\otimes D_{2}:G1\times G2\rightarrow{\rm End}(V1\otimes V2)$,
defined for any couple $(g1,g2)\in G1\times G2$ by $D_{\otimes}(g1,g2)=D1(g1)\otimes D2(g2)$
is again irreducible. Indeed the Schur orthogonality relation for
an irreducible representation states (with normalized measure with
respect to the group volume) that : 
\[
\int d\eta(g)|\chi(g)|^{2}=1
\]
where $\chi(g)$ is the character of the representation. Then : 
\begin{eqnarray*}
\int d\eta(g_{1}\times g_{2})|\chi_{\otimes}(g_{1}\times g_{2})|^{2} & = & \int d\eta(g_{1})|\chi_{1}(g_{1})|^{2}\int d\eta(g_{2})|\chi_{2}(g_{2})|^{2}\\
 & = & 1
\end{eqnarray*}
and the representation $D_{\otimes}$ is again irreducible. 

By Schur lemma, we then have that $\mathbb{E}[(\rho_{0}^{\otimes2})^{(i_{1}i_{2}i_{1}i_{2})}]\propto\mathbb{I}^{(i_{1}i_{2}i_{1}i_{2})}$
where $\mathbb{I}^{(i_{1}i_{2}i_{1}i_{2})}$ is the identity defined
by : $\mathbb{I}^{(i,j,k,l)}\equiv\sum_{\nu_{i},\nu_{j},\nu_{k},\nu_{l}}\ket{E_{i}^{\nu_{i}},E_{j}^{\nu_{j}}}\bra{E_{k}^{\nu_{k}},E_{l}^{\nu_{l}}}.$
As before, we take the trace to determine the proportionality coefficient,
we get : 
\[
\mathbb{E}[(\rho_{0}^{\otimes2})^{(i_{1}i_{2}i_{1}i_{2})}]=\frac{\mathrm{tr}(\rho_{0}^{(i_{1}i_{1})})\mathrm{tr}(\rho_{0}^{(i_{2}i_{2})})}{d_{i_{1}}d_{i_{2}}}\mathbb{I}^{(i_{1}i_{2}i_{1}i_{2})}.
\]

\paragraph{Case III : }

We look at : 

\[
\mathbb{E}[(\rho_{0}^{\otimes2})^{(i_{1}i_{2}i_{2}i_{1})}]=\int d\eta(U)U^{(i_{1})}\otimes U^{(i_{2})}(\rho_{0}^{\otimes2})^{(i_{1}i_{2}i_{2}i_{1})}U^{\dagger(i_{2})}\otimes U^{\dagger(i_{1})}.
\]
Let $M$ be an element of $M\in L({\cal H}_{i},{\cal H}_{j})\otimes L({\cal H}_{k},{\cal H}_{l})$
and $\sigma\in\mathfrak{S}_{2}$. We define the \emph{right action
}of $\sigma$ on $M$, $M\cdot\sigma$ as : 
\[
(M\cdot\sigma)_{\alpha_{1}\beta_{1}\alpha_{2}\beta_{2}}=M_{\alpha_{1}\beta_{\sigma(1)}\alpha_{2}\beta_{\sigma(2)}}
\]
As it will be useful later, we define in
the same way, the \emph{left action} $\sigma\cdot M$ acting on $M$
as : 
\[
(\sigma\cdot M)_{\alpha_{1}\beta_{1}\alpha_{2}\beta_{2}}=M_{\alpha_{\sigma(1)}\beta_{1}\alpha_{\sigma(2)}\beta_{2}}
\]

We then have that : 
\begin{eqnarray*}
\mathbb{E}[(\rho_{0}^{\otimes2})^{(i_{1}i_{2}i_{2}i_{1})}] & = & \int d\eta(U)U^{(i_{1})}\otimes U^{(i_{2})}(\rho_{0}^{\otimes2})^{(i_{1}i_{2}i_{2}i_{1})}U^{\dagger(i_{\sigma(1)})}\otimes U^{\dagger(i_{\sigma(2)})}\\
\mathbb{E}[(\rho_{0}^{\otimes2})^{(i_{1}i_{2}i_{2}i_{1})}]\cdot\sigma & = & (\int d\eta(U)U^{(i_{1})}\otimes U^{(i_{2})}((\rho_{0}^{\otimes2})^{(i_{1}i_{2}i_{2}i_{1})}.\sigma)U^{\dagger(i_{1})}\otimes U^{\dagger(i_{2})})
\end{eqnarray*}
where $\sigma$ is here the permutation $\{1,2\}\to\{2,1\}$.

Applying Schur lemma as before then leads to : 
\begin{eqnarray*}
\mathbb{E}[(\rho_{0}^{\otimes2})^{(i_{1}i_{2}i_{2}i_{1})}]\cdot\sigma & = & \frac{\mathrm{tr}((\rho_{0}^{\otimes2})^{(i_{1}i_{2}i_{2}i_{1})}.\sigma)}{d_{i_{1}}d_{i_{2}}}\mathbb{I}^{(i_{1}i_{1}i_{2}i_{2})},\\
\mathbb{E}[(\rho_{0}^{\otimes2})^{(i_{1}i_{2}i_{2}i_{1})}] & = & \frac{\mathrm{tr}(\rho_{0}^{(i_{1}i_{2})}\rho_{0}^{(i_{2}i_{1})})}{d_{i_{1}}d_{i_{2}}}\mathbb{I}^{(i_{1}i_{2}i_{2}i_{1})},\\
 & = & \frac{\mathrm{tr}((\rho_{0}^{\otimes2})\mathbb{I}^{(i_{1}i_{2}i_{2}i_{1})})}{d_{i_{1}}d_{i_{2}}}\mathbb{I}^{(i_{1}i_{2}i_{2}i_{1})}.
\end{eqnarray*}
Regrouping the results for all three cases proves (\ref{eq:moment2})
of the main text :

\begin{eqnarray*}
\mathbb{E}[\rho_{0}^{\otimes2}] & = & \sum_{i_{1}}\frac{1}{d_{i_{1}}(d_{i_{1}}+1)}(({\rm tr}(\rho_{0}^{(i_{1})}))^{2}+{\rm tr}((\rho_{0}^{(i_{1})})^{2}))\mathbb{I}_{(2)}^{(i_{1},i_{1},i_{1},i_{1})}\\
 &  & +\frac{1}{d_{i_{1}}(d_{i_{1}}-1)}(({\rm tr}(\rho_{0}^{(i_{1})}))^{2}-{\rm tr}((\rho_{0}^{(i_{1})})^{2}))\mathbb{I}_{(1,1)}^{(i_{1},i_{1},i_{1},i_{1})}\\
 &  & +\sum_{i_{1}\neq i_{2}}\frac{1}{d_{i_{1}}d_{i_{2}}}(\mathrm{tr}(\rho_{0}^{(i_{1}i_{1})})\mathrm{tr}(\rho_{0}^{(i_{2}i_{2})})\mathbb{I}^{(i_{1}i_{2}i_{1}i_{2})}+\mathrm{tr}(\rho_{0}^{(i_{1}i_{2})}\rho_{0}^{(i_{2}i_{1})})\mathbb{I}^{(i_{1}i_{2}i_{2}i_{1})})
\end{eqnarray*}
\section{General formula at any order \label{sec:General-formula-at}}
In this section, we wish to compute the generalization of the formula
for the mean and second order correlation of density matrix elements
to higher order, i.e :

\begin{eqnarray}
\mathbb{E}[\rho_{0}^{\otimes n}] & = & \int_{{\cal G}}d\eta(U)\mathrm{Tr}(U^{\otimes n}\rho_{0}^{\otimes n}U^{\dagger\otimes n})\label{eq:highermomenta}
\end{eqnarray}
This is equivalent to knowing the generating
function $Z(A)$ defined as
\[
Z(A)\equiv\int_{{\cal G}}d\eta(U)e^{{\rm tr}(AU\rho_{0}U^{\dagger})}
\]
which is reminiscent of the Harish-Chandra Itzykson Zuber integral
\cite{planarapproximationII} except that the group ${\cal G}$ upon
which the integration is performed is \emph{not }an unitary group
so we can't directly use that result.

To compute (\ref{eq:highermomenta}), we will rely on the same approach
than for the mean and the second order correlations, i.e we will first
decompose $\rho_{0}^{\otimes n}$ into different \emph{sectors }transforming
according to different combination of $U^{(i_{k})}$ and identify
the different \emph{invariants }under such transformations. In spirit,
it will be close to the proof of the invariant theory presented in
the appendix of \cite{4BBJ}. We introduce once again the block decomposition
of $\rho_{0}^{\otimes n}$ into $(\rho_{0}^{\otimes n}){}^{(\boldsymbol{i}\boldsymbol{j})}\equiv(\rho_{0}^{\otimes n}){}^{(i_{1}i_{2}\cdots i_{n},j_{1}j_{2}\cdots j_{n})}$
defined by : 
\begin{align*}
 & (\rho_{0}^{\otimes n})^{(i_{1}i_{2}\cdots i_{n},j_{1}j_{2}\cdots j_{n})}\\
 & =\sum_{\nu_{i_{1}},\cdots,\nu_{i_{n}},\nu_{j_{1}},\cdots,\nu_{n}}{\rm tr}(\rho_{0}^{\otimes n}\ket{E_{i_{1}}^{\nu_{i_{1}}},\cdots,E_{i_{n}}^{\nu_{i_{n}}}}\bra{E_{j_{1}}^{\nu_{j_{1}}},\cdots,E_{j_{n}}^{\nu_{j_{n}}}})\ket{E_{i_{1}}^{\nu_{i_{1}}},\cdots,E_{i_{n}}^{\nu_{i_{n}}}}\bra{E_{j_{1}}^{\nu_{j_{1}}},\cdots,E_{j_{n}}^{\nu_{j_{n}}}}
\end{align*}
The average of a block is given by : 
\[
\mathbb{E}[(\rho_{0}^{\otimes n}){}^{(\boldsymbol{i}\boldsymbol{j})}]=\int_{U\in{\cal G}}d\eta(U)U^{(i_{1})}\otimes U^{(i_{2})}\cdots\otimes U^{(i_{n})}(\rho_{0}^{\otimes n}){}^{(\boldsymbol{i}\boldsymbol{j})}U^{\dagger(j_{1})}\otimes U^{\dagger(j_{2})}\cdots\otimes U^{\dagger(j_{n})}
\]
To each term $U^{(i_{1})}\otimes U^{(i_{2})}\cdots\otimes U^{(i_{n})}$,
we associate a \emph{standard ordering }defined as the tensor product
\begin{eqnarray*}
U_{{\rm so}}^{(m_{1}\cdots m_{k_{{\rm max}}})} & \equiv & U^{(i_{\sigma^{-1}(1)})}\otimes U^{(i_{\sigma^{-1}(2)})}\cdots\otimes U^{(i_{\sigma^{-1}(n)})}\\
 & \equiv & U^{(m_{1})\otimes n_{1}}\otimes\cdots\otimes U^{(m_{n})\otimes n_{k}}.
\end{eqnarray*}
 where $i_{\sigma^{-1}(1)}\leq i_{\sigma^{-1}(2)}\cdots\leq i_{\sigma^{-1}(n)}$,
$\sigma\in\frac{\mathfrak{S}_{n}}{\prod\mathfrak{\mathfrak{S}}_{n_{k}}}$,
$m_{k}=i_{\sigma^{-1}(\sum_{j=1}^{k}n_{j})}$ and $n_{k}$ is the
number of times $U^{(m_{k})}$ appears in the tensor product (we have
$\sum_{k=1}^{k_{{\rm max}}}n_{k}=n$). 

In the same way, we define the permutation $\sigma^{*}\in\frac{\mathfrak{S}_{n}}{\prod\mathfrak{\mathfrak{S}}_{n_{k}}}$
such that $U^{\dagger(j{}_{\sigma^{*-1}(1)})}\otimes U^{\dagger(j_{\sigma^{*-1}(2)})}\cdots\otimes U^{\dagger(j_{\sigma^{*-1}(n)})}$
is standard ordered.

We can use the same argument as before to show that the average is
null unless the $j_{k}'s$ are a permutation of the $i_{k}$'s. We
call this permutation $\gamma$ : $j_{\gamma(k)}=i_{k}$. $\gamma$
is related to $\sigma$, $\sigma^{*}$ by $\gamma=\sigma^{*-1}\sigma$.
Indeed : 
\begin{eqnarray*}
i_{\sigma^{-1}(k)} & = & j_{\sigma^{*-1}(k)}\\
i_{k} & = & j_{\sigma^{*-1}(\sigma(k))}\\
j_{\gamma(k)} & = & j_{\sigma^{*-1}(\sigma(k))}.
\end{eqnarray*}
Now :
\begin{eqnarray*}
\mathbb{E}[(\rho_{0}^{\otimes n}){}^{(\boldsymbol{i}\boldsymbol{j})}] & = & \int_{U\in{\cal G}}d\eta(U)U^{(i_{\sigma(\sigma^{-1}(1))})}\cdots\otimes U^{(i_{\sigma(\sigma^{-1}(n))})}(\rho_{0}^{\otimes n}){}^{(\boldsymbol{i}\boldsymbol{j})}U^{(j_{\sigma^{*}(\sigma^{*-1}(1))})}\cdots\otimes U^{\dagger(j_{\sigma^{*}(\sigma^{*-1}(n))})}\\
\sigma\cdot\mathbb{E}[(\rho_{0}^{\otimes n}){}^{(\boldsymbol{i}\boldsymbol{j})}] & \cdot\sigma^{*}= & \int_{U\in{\cal G}}d\eta(U)U_{{\rm so}}^{(m_{1},\cdots,m_{k_{{\rm max}}})}\sigma\cdot(\rho_{0}^{\otimes n}){}^{(\boldsymbol{i}\boldsymbol{j})}\cdot\sigma^{*}U_{{\rm so}}^{\dagger(m_{1},\cdots,m_{k_{{\rm max}}})}.
\end{eqnarray*}
In general, the tensor product representation which $U_{{\rm so}}^{(m_{1},\cdots,m_{k_{{\rm max}}})}$
belongs to is reducible : 
\begin{eqnarray*}
U_{{\rm so}}^{(m_{1},\cdots,m_{k_{{\rm max}}})} & = & U^{(m_{1})\otimes n_{1}}\otimes\cdots\otimes U^{(m_{k_{{\rm max}}})\otimes n_{k_{{\rm max}}}}.\\
 & = & \oplus_{(y_{1})\in\{n_{1}\}}D^{(y_{1})}(U^{(m_{1})})\otimes\oplus_{(y_{2})\in\{n_{2}\}}D^{(y_{2})}(U^{(m_{2})})\cdots\otimes\oplus_{y_{k_{{\rm max}}}\{n_{k_{{\rm max}}}\}}D^{(y_{k_{{\rm max}}})}(U^{(m_{k_{{\rm max}}})})\\
 & = & \oplus_{\{n_{1}\},\{n_{2}\},\cdots\{n_{k_{{\rm max}}}\}}D{}^{(y_{1})}(U^{(m_{1})})\otimes\cdots\otimes D^{(y_{k_{{\rm max}}})}(U^{(m_{k_{{\rm max}}})})\\
 & \equiv & \oplus_{\{\boldsymbol{n}\}}\otimes_{i}D^{(y_{i})}(U^{(m_{i})})
\end{eqnarray*}
where as before $\{n_{k}\}$ designates the possible Young tableaux
of $n_{k}$. As we showed before, the tensor product of two irreducible
representations is again irreducible, so the representations $D^{(y_{1})}\otimes\cdots\otimes D^{(y_{k_{{\rm max}}})}$
of $\times_{i}SU(d_{i})^{n_{k}}$ are \emph{irreducible}. As before,
we decompose further $\sigma\cdot(\rho_{0}^{\otimes n}){}^{(\boldsymbol{i}\boldsymbol{j})}.\sigma^{*}$
into blocks corresponding to these irreducible representations. Denoting
by $\ket{E_{i},(y),\nu}$ the basis elements associated to the irreducible
representation of $SU(d_{i})^{n}$ corresponding to the decomposition
$(y)$, we have the following decomposition for the blocks :
\begin{align*}
 & (\sigma.(\rho_{0}^{\otimes n}){}^{(\boldsymbol{i},\boldsymbol{j})}.\sigma^{*})^{(\{\boldsymbol{y}\},\{\boldsymbol{y'}\})}\\
 & =\sum_{\nu_{1},\cdots,\nu_{k_{{\rm max}}}\nu'_{1},\cdots,\nu'_{k_{{\rm max}}}}\\
 & {\rm tr((\sigma\cdot(\rho_{0}^{\otimes n}){}^{(\boldsymbol{i}\boldsymbol{j})}.\sigma^{*})\left|E_{m_{1}},(y_{1}),\nu_{1},\cdots,E_{m_{k_{{\rm max}}}},(y_{k_{{\rm max}}}),\nu_{k_{{\rm max}}}\right\rangle \left\langle E_{m_{1}},(y{}_{1}^{'}),\nu'_{1},\cdots,E_{m_{k_{{\rm max}}}},(y'_{k_{{\rm max}}}),\nu'_{k_{{\rm max}}}\right|)}\\
 & \left|E_{m_{1}},(y_{1}),\nu_{1},\cdots,E_{m_{k_{{\rm max}}}},(y_{k_{{\rm max}}}),\nu_{k_{{\rm max}}}\right\rangle \left\langle E_{m_{1}},(y{}_{1}^{'}),\nu'_{1},\cdots,E_{m_{k_{{\rm max}}}},(y'_{k_{{\rm max}}}),\nu'_{k_{{\rm max}}}\right|
\end{align*}
where $\{\boldsymbol{y}\}$ designates the tuple $\{y_{j}\}_{1\leq j\leq k_{{\rm max}}}$.
The averages of these blocks are given by 
\begin{eqnarray*}
 &  & \mathbb{E}[(\sigma.(\rho_{0}^{\otimes n}){}^{(\boldsymbol{i},\boldsymbol{j})}.\sigma^{*})^{(\{\boldsymbol{y}\},\{\boldsymbol{y'}\})}]\\
 &  & =\int_{{\cal G}}D^{(y_{1})}(U^{(m_{1})})\otimes\cdots\otimes D^{(y_{k_{{\rm max}}})}(U^{(m_{k_{{\rm max}}})})(\sigma.(\rho_{0}^{\otimes n}){}^{(\boldsymbol{i},\boldsymbol{j})}.\sigma^{*})^{(\{\boldsymbol{y}\},\{\boldsymbol{y'}\})}\\
 &  & D^{\dagger(y'_{1})}(U^{(m_{1}')})\otimes\cdots\otimes D^{\dagger(y'_{k_{{\rm max}}})}(U^{(m'_{k_{{\rm max}}})}).
\end{eqnarray*}
As before, by Schur lemma, the average of one of these blocks is non
zero only if the two representations are equivalent, meaning that
we must have $\{\boldsymbol{y}\}=\{\boldsymbol{y}'\}$. Then :
\begin{eqnarray*}
\mathbb{E}[(\sigma.(\rho_{0}^{\otimes n}){}^{(\boldsymbol{i},\boldsymbol{j})}.\sigma^{*})^{(\{\boldsymbol{y}\},\{\boldsymbol{y'}\})}] & = & \delta_{\boldsymbol{y},\boldsymbol{y'}}\prod_{k}\frac{\omega((y_{k}))}{d_{(y_{k})}}\mathrm{tr}((\sigma\cdot(\rho_{0}^{\otimes n}){}^{(\boldsymbol{i}\boldsymbol{j})}\cdot\sigma^{*})^{(\{\boldsymbol{y}\},\{\boldsymbol{y}\})})(\sigma.\mathbb{I}^{(\boldsymbol{i},\boldsymbol{j})}.\sigma^{*})^{(\{\boldsymbol{y}\},\{\boldsymbol{y}\})},
\end{eqnarray*}
where $\omega((y_{k}))$ is the multiplicity of the Young tableau
corresponding to $(y_{k})$, $d_{(y_{k})}$ its dimension and $(\sigma.\mathbb{I}^{(\boldsymbol{i}\boldsymbol{j})}.\sigma^{*})^{(\{\boldsymbol{y}\}\{\boldsymbol{y}\})}$
is defined as : 
\begin{align*}
 & (\sigma.\mathbb{I}^{(\boldsymbol{i}\boldsymbol{j})}.\sigma^{*})^{(\{\boldsymbol{y}\}\{\boldsymbol{y}\})}\\
 & \equiv\sum_{\nu_{1},\cdots\nu_{k_{{\rm max}}}}\left|E_{m_{1}},(y_{1}),\nu_{1},\cdots,E_{m_{k_{{\rm max}}}},(y_{k_{{\rm max}}}),\nu_{k_{{\rm max}}}\right\rangle \left\langle E_{m_{1}},(y_{1}),\nu_{1},\cdots,E_{m_{k_{{\rm max}}}},(y_{k_{{\rm max}}}),\nu_{k_{{\rm max}}}\right|.
\end{align*}
This finally leads us to : 
\[
\mathbb{E}[\rho_{0}^{\otimes n}]=\sum_{(\boldsymbol{i},\boldsymbol{j})}\sum_{(\{\boldsymbol{y}\})}\prod_{k}\frac{\omega((y_{k}))}{d_{(y_{k})}}\mathrm{tr}((\sigma\cdot(\rho_{0}^{\otimes n}){}^{(\boldsymbol{i}\boldsymbol{j})}\cdot\sigma^{*})^{(\{\boldsymbol{y}\},\{\boldsymbol{y}\})})\mathbb{I}^{(\boldsymbol{i}\boldsymbol{j})}{}^{(\{\boldsymbol{y}\},\{\boldsymbol{y}\})}.
\]
where the sum is over all $\boldsymbol{j}$'s that are a permutation
of $\boldsymbol{i}$'s. Each permutation is characterized by $\sigma$
and $\sigma^{*}$.
\section{More deta\label{sec:More-details-on}ils on the case study}

\begin{figure}[H]
\begin{centering}
\includegraphics[scale=0.4]{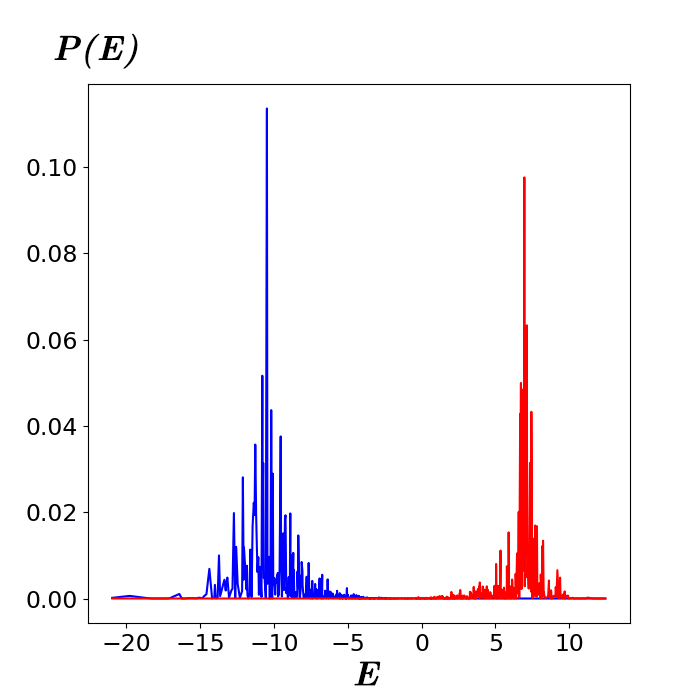}
\par\end{centering}
\caption{Decomposition of $\ket{\Phi_{1}}$ (blue) and $\ket{\Phi_{2}}$ (red)
in the eigenbasis of $\hat{H}$. \label{fig:Decomposition-of-}}
\end{figure}

In this appendix, we provide additional details on the numerics presented
in the main text. As a reminder, the Hamiltonian we study is :

\[
\hat{H}=\sum_{j=0}^{L-2}J\vec{\sigma}_{j}\cdot\vec{\sigma}_{j+1}+\sum_{j=0}^{L-1}h_{j}\sigma_{j}^{z}
\]
with $L=12$, $J=1$ and $h_{j}$ picked at random between $-1$ and
$1$ with the uniform distribution. We work in the $0$ magnetization
sector which has a dimension of $924$. We choose a seed such that
the mean level spacing of the spectrum is close to the Wigner distribution
one : $0.53069$. The minimum energy $E_{{\rm min}}$ is $-20.944$
and the maximum energy $E_{{\rm max}}$ is $12.445$. The states $\ket{\Phi_{1}}$
and $\ket{\Phi_{2}}$ have respectively energies $E_{1}\equiv-10.753$
and $E_{2}\equiv6.731$ with respect to $\hat{H}_{{\rm R}}+\hat{H}_{{\rm L}}$.
One important quantity is the overlap $O_{{\rm vlap}}$ these states
have with respect to the eigenbasis $\ket{i}$ of the total Hamiltonian,
i.e $O_{{\rm vlap}}(i)\equiv|\braket{i|\Phi_{1}}\braket{i|\Phi_{2}}|$.
The maximum of $O_{{\rm vlap}}$ in our case is $8.206*10^{-5}$.
We also have that $\sum_{i}O_{{\rm vlap}}(i)=0.00579$. The decompositions
of $\left|\Phi_{1}\right\rangle $ and $\left|\Phi_{2}\right\rangle $
in the eigenbasis of $\hat{H}$ are shown on fig.\ref{fig:Decomposition-of-}.

To compute the numerical values, we choose a time window {[}3000 :
13000{]} with $2*10^{4}$ points. The different values presented in
tab.\ref{tab:} of the main text are given by time average over this
interval : 
\begin{align*}
\mathbb{E}_{t}[A(t)] & =\frac{1}{T}\int_{0}^{T}dtA(t)\\
\sigma_{A} & \equiv\sqrt{\frac{1}{T}\int_{0}^{T}(A(t)-\mathbb{E}_{t}[A(t)])^{2}}
\end{align*}
with $T$ the time interval. 

The confidence intervals $\delta$ on the mean and the standard deviation are obtained
by dividing $T$ in 10 smaller intervals on which the quantities of
interest are computed again. $\delta$ then corresponds to the standard
deviation between the results obtained on the 10 samples and the one
computed on the full interval. 

The theoretical values for the mean and variances of our different
quantities are computed from equations (\ref{eq:mean}) and (\ref{eq:moment2})
of the main text. Recall that for the protocol I, the initial state
was chosen as the pure state $\frac{1}{\sqrt{2}}(\ket{\Phi_{1}}+\ket{\Phi_{2}})$
and for the protocol II, it was the classical mixture described by
the density matrix $\rho_{0}=\frac{1}{2}(\ket{\Phi_{1}}\bra{\Phi_{1}}+\ket{\Phi_{2}}\bra{\Phi_{2}})$. 

For a fully degenerate spectrum, (\ref{eq:mean},\ref{eq:moment2})
gives that the average of a given observable $\hat{M}$ and its second
order connected correlations are given by : 
\begin{align*}
{\rm tr}(\mathbb{E}[\rho_{0}]\hat{M})= & \sum_{i}\left\langle i\left|\rho_{0}\right|i\right\rangle \left\langle i\right|\hat{M}\left|i\right\rangle \\
{\rm tr}(\mathbb{E}[\rho_{0}^{\otimes2}]\hat{M}\otimes\hat{M})^{{\rm c}}= & \sum_{i_{1}\neq i_{2}}|\left\langle i_{1}\left|\rho_{0}\right|i_{2}\right\rangle |^{2}|\left\langle i_{1}\right|\hat{M}\left|i_{2}\right\rangle |^{2}
\end{align*}
Below, we explain why there is no difference in the first and second
order correlation of $\hat{H}_{{\rm R}}$ between protocol I and II
while there are for the second order correlation of $\hat{Q}$.

\paragraph{Mean quantities :}

\begin{eqnarray}
{\rm tr}(\mathbb{E}[\rho_{0}^{({\rm I})}]\hat{H}_{{\rm R}}) & = & \sum_{i}\left\langle i\left|\rho_{0}\right|i\right\rangle \left\langle i\right|\hat{H}_{{\rm R}}\left|i\right\rangle \nonumber \\
 & = & \frac{1}{2}\sum_{i}\bra{i}(\ket{\Phi_{1}}+\ket{\Phi_{2}})(\bra{\Phi_{1}}+\bra{\Phi_{2}})\ket{i}\left\langle i\right|\hat{H}_{{\rm R}}\left|i\right\rangle 
\end{eqnarray}
Since the overlap between $\ket{\Phi_{1}}$ and $\ket{\Phi_{2}}$
is small, we can approximate the previous expression by : 
\begin{eqnarray*}
{\rm tr}(\mathbb{E}[\rho_{0}^{({\rm I})}]\hat{H}_{{\rm R}}) & \approx & \frac{1}{2}\sum_{i}\bra{i}\hat{H}_{{\rm R}}\ket{i}\bra{i}(\ket{\Phi_{1}}\bra{\Phi_{1}}+\ket{\Phi_{2}}\bra{\Phi_{2}})\ket{i}\\
 & = & {\rm tr}(\mathbb{E}[\rho_{0}^{({\rm II})}]\hat{H}_{{\rm R}})
\end{eqnarray*}
Similarly :
\begin{eqnarray*}
{\rm tr}(\mathbb{E}[\rho_{0}^{({\rm I})}]\hat{Q}) & = & \sum_{i}\braket{i|\Phi_{1}}\braket{\Phi_{2}|i}\bra{i}\rho_{0}^{({\rm I})}\ket{i}\\
 & \approx & 0\\
 & = & {\rm tr}(\mathbb{E}[\rho_{0}^{({\rm II})}]\hat{Q})
\end{eqnarray*}
We see that as far as the mean quantities are concerned, the cat state
and the classical mixture provide the same results. 

\paragraph{Second order fluctuations :}

From (\ref{eq:moment2}), we have that : 
\begin{eqnarray*}
 &  & {\rm tr}(\mathbb{E}[\rho_{0}^{({\rm I})\otimes2}]\hat{H}_{{\rm R}}\otimes\hat{H}_{{\rm R}})^{{\rm c}}\\
 &  & =\sum_{i_{1}\neq i_{2}}|\left\langle i_{1}\right|\rho_{0}^{({\rm I})}\left|i_{2}\right\rangle |^{2}\left\langle i_{1}\right|\hat{H}_{{\rm R}}\left|i_{2}\right\rangle |^{2}\\
 &  & =\frac{1}{4}\sum_{i_{1}\neq i_{2}}|\left\langle i_{1}\left|(\left|\Phi_{1}\right\rangle \left\langle \Phi_{1}\right|+\left|\Phi_{1}\right\rangle \left\langle \Phi_{2}\right|+\left|\Phi_{2}\right\rangle \left\langle \Phi_{1}\right|+\left|\Phi_{2}\right\rangle \left\langle \Phi_{2}\right|)\right|i_{2}\right\rangle |^{2}|\left\langle i_{1}\right|\hat{H}_{{\rm R}}\left|i_{2}\right\rangle |^{2}\\
 &  & \approx\frac{1}{4}\sum_{i_{1}\neq i_{2}}|\left\langle i_{1}\left|(\left|\Phi_{1}\right\rangle \left\langle \Phi_{1}\right|+\left|\Phi_{2}\right\rangle \left\langle \Phi_{2}\right|)\right|i_{2}\right\rangle |^{2}|\left\langle i_{1}\right|\hat{H}_{{\rm R}}\left|i_{2}\right\rangle |^{2}\\
 &  & \approx{\rm tr}(\mathbb{E}[\rho_{0}^{({\rm II})\otimes2}]\hat{H}_{{\rm R}}\otimes\hat{H}_{{\rm R}})^{{\rm c}}
\end{eqnarray*}
Where in the last line we used the fact that $\left\langle i_{1}|\Phi_{1}\right\rangle \left\langle \Phi_{2}|i_{2}\right\rangle $
is non zero only for $E_{i_{1}}$ close to $E_{1}$ and $E_{i_{2}}$
close to $E_{2}$. But this in turns imply that $\left\langle i_{1}\right|\hat{H}_{{\rm R}}\left|i_{2}\right\rangle \approx0$
.

For $\hat{Q}$ : 
\begin{align*}
 & {\rm tr}(\mathbb{E}[\rho_{0}^{({\rm I})\otimes2}]\hat{Q}\otimes\hat{Q})^{{\rm c}}\\
 & =\frac{1}{4}\sum_{i_{1}\neq i_{2}}|\left\langle i_{1}\left|(\left|\Phi_{1}\right\rangle \left\langle \Phi_{1}\right|+\left|\Phi_{1}\right\rangle \left\langle \Phi_{2}\right|+\left|\Phi_{2}\right\rangle \left\langle \Phi_{1}\right|+\left|\Phi_{2}\right\rangle \left\langle \Phi_{2}\right|)\right|i_{2}\right\rangle |^{2}|\left\langle i_{1}\right|\hat{Q}\left|i_{2}\right\rangle |^{2}\\
 & =\frac{1}{4}\sum_{i_{1}\neq i_{2}}|\left\langle i_{1}\left|(\left|\Phi_{1}\right\rangle \left\langle \Phi_{1}\right|+\left|\Phi_{1}\right\rangle \left\langle \Phi_{2}\right|+\left|\Phi_{2}\right\rangle \left\langle \Phi_{1}\right|+\left|\Phi_{2}\right\rangle \left\langle \Phi_{2}\right|)\right|i_{2}\right\rangle |^{2}|\left\langle i_{1}|\Phi_{1}\right\rangle \left\langle \Phi_{2}|i_{2}\right\rangle +\left\langle i_{1}|\Phi_{2}\right\rangle \left\langle \Phi_{1}|i_{2}\right\rangle |^{2}\\
 & \approx\frac{1}{4}\sum_{i_{1}\neq i_{2}}|\left\langle i_{1}|\Phi_{1}\right\rangle \left\langle \Phi_{2}|i_{2}\right\rangle +\left\langle i_{1}|\Phi_{2}\right\rangle \left\langle \Phi_{1}|i_{2}\right\rangle |^{4}
\end{align*}
Where to go to the last line we used the fact that if $|\left\langle i_{1}|\Phi_{1}\right\rangle \left\langle \Phi_{2}|i_{2}\right\rangle +\left\langle i_{1}|\Phi_{2}\right\rangle \left\langle \Phi_{1}|i_{2}\right\rangle |^{2}$
is non zero, the $E_{i_{1}}$ is close to either $E_{1}$ or $E_{2}$
and $E_{i_{2}}$ the other way around. This in turn implies that $\left\langle i_{1}|\Phi_{1}\right\rangle \left\langle \Phi_{1}|i_{2}\right\rangle \approx0\approx\left\langle i_{1}|\Phi_{2}\right\rangle \left\langle \Phi_{2}|i_{2}\right\rangle $.
This also tells us that 
\[
{\rm tr(\mathbb{E}[\rho_{0}^{({\rm II})\otimes2}\hat{Q}\otimes\hat{Q}]\approx0}
\]
We thus see a clear difference in the fluctuations of $\hat{Q}$ for
the two protocols. 

\end{appendix}



\bibliography{bibliotony.bib}

\nolinenumbers

\end{document}